\documentclass[fleqn,10pt,a4paper]{article}
\usepackage{amsmath,amssymb,bm}
\usepackage{cite,epsfig,color,graphicx}

\title{Testing vacuum electrodynamics using `slow light' experiments} 

\author{S P Flood\thanks{Department of Physics,
Lancaster University, Lancaster, LA1 4YB, UK
and Cockcroft Institute, Daresbury, WA4 4AD, UK}
\and
D A Burton\footnotemark[1]
}
\begin{document}
\maketitle
\begin{abstract}
A recent proposal to explore vacuum electrodynamics using the speed of propagation of an electromagnetic pulse through an ambient constant magnetic field is examined. It is argued that the proposal should be modified so that the background magnetic field, the direction of propagation and the transverse projection of the electric field (with respect to the direction of propagation) are not coplanar. The implications of invariance under Gibbons' electric-magnetic duality rotations are determined in this context.
\end{abstract}
\section{Introduction}
The scattering of light by light in the vacuum has been a source of considerable interest over many decades. The most common approach to modelling electrodynamics in high-field scenarios employs effective actions extracted from the QED vacuum-to-vacuum persistence amplitude of the electron-positron field in an ambient electromagnetic field~\cite{dittrich:2000}. Although the QED vacuum is expected to feature prominently in non-terrestrial strong field environments, such as gamma-ray pulsars and magnetars~\cite{lai:2003}, the study of non-linear electrodynamics has received considerable impetus from advances in terrestrial high-power laser facilities in recent years~\cite{marklund:2010, king:2010}. In particular, the anticipated laser field intensities in the forthcoming Extreme Light Infrastructure \cite{eli} are so large that photon-photon scattering will be appreciable.

One may speculate that the Standard Model quantum vacuum is not the only source of an effective self-coupling of the electromagnetic field. In particular, the partition functional for an open string coupled to an ambient $U(1)$ gauge potential leads to the Born-Infeld effective action~\cite{fradkin:1985} for the electromagnetic field and, from an aesthetic perspective, Born-Infeld electrodynamics~\cite{born:1934} is privileged because it is unique in retaining the exceptional causal properties of Maxwell theory~\cite{boillat:1970}. More precisely, it is the only regular generalization of Maxwell electrodynamics generated from an arbitrary local Lagrangian depending only on the two electromagnetic invariants that also has zero birefringence and does not exhibit shock waves. Although the value of the coupling constant governing the self-interaction of the electromagnetic field in Euler-Heisenberg electrodynamics is determined by the electron rest mass and charge, the analogous quantity in string-inspired Born-Infeld electrodynamics (string tension) is unknown. Moreover, string-theoretic and other considerations have been used to motivate a range of ``Born-Infeld-like'' theories (e.g. ~\cite{ayon-beato:1999, kruglov:2010, burton:2011b}) and a unique extension to classical Maxwell theory does not immediately present itself in the wider context.

Recent investigations of the implications of non-linear electrodynamics~\cite{burton:2011a, burton:2011b, tucker:2010, ferraro:2010, kruglov:2010, munoz:2009, aiello:2007, denisov:2000} have concentrated on properties of the electromagnetic field within particular theories. However, little attention has been paid thus far to {\it indistinguishable} aspects of different theories and attendant implications for proposed tests of non-linear electrodynamics.

Exact Born-Infeld waves propagating in a constant background (electric or magnetic) field on flat spacetime were explored in Ref.~\cite{aiello:2007}, and the phase speed of the wave was found to depend on the ambient background field. A similar `slow light' effect was reported in Ref.~\cite{tucker:2010} where the transit time of an electromagnetic pulse across a region bathed in a constant magnetic field was examined within Born-Infeld electrodynamics. In Ref.~\cite{tucker:2010} the electric field of the pulse and ambient magnetic field were aligned; in the following we argue that such an experiment would be more effective if a more general configuration is employed.

This paper uses the Einstein summation convention throughout. Latin indices run from 0 to 3 and units are used in which the speed of light $c=1$ and the permittivity of free space $\epsilon_0=1$. 

The following work is formulated on a flat spacetime manifold with frame $\{X_a =\partial/\partial x^a\}$, naturally dual co-frame $\{e^a=dx^a\}$ and metric $g$ given by
\begin{equation}
g = -e^0\otimes e^0 +\sum^{3}_{i=1}e^i\otimes e^i.
\end{equation}
The spacetime volume element is
\begin{equation}
\star 1 = dx^0\wedge dx^1\wedge dx^2\wedge dx^3
\end{equation}
where $\star$ denotes the Hodge map and $\{x^a\}$ denotes $\{t,x,y,z\}$ as the standard coordinate system for the lab frame. Then the macroscopic equations describing the electromagnetic field in the vacuum are given by
\begin{align}
&dF = 0, \label{maxwell1}\\
&d\star G = 0\label{maxwell2}.
\end{align}
Here $F$ is the Faraday 2-form
\begin{equation}
F = dt\wedge E + \star (dt\wedge B)
\end{equation}
(encapsulating information about the electric field and magnetic field) and $G$ is the excitation 2-form related to $F$ by the expression
\begin{equation}
G = 2\left(\frac{\partial\mathcal{L}}{\partial X}F-\frac{\partial\mathcal{L}}{\partial Y}\star F\right)\label{Gdef}
\end{equation}
where the local $0$-form $\mathcal{L}(X,Y)$ is the Lagrangian of the electromagnetic theory and is assumed to depend only on the electromagnetic invariants $X$ and $Y$,
\begin{align}
\label{X_def}
&X = \star(F\wedge\star F),\\
\label{Y_def}
&Y = \star(F\wedge F).
\end{align}
For example, the Lagrangians for linear Maxwell theory and vacuum Born-Infeld electrodynamics are 
\begin{align}
&\mathcal{L}_{\rm M}=\frac{X}{2},\\
&\mathcal{L}_{\rm BI}=\frac{1}{\kappa^2}\left(1-\sqrt{1-\kappa^2X -\frac{\kappa^4}{4}Y^2}\right)\label{BI_def}
\end{align}
respectively, where $\kappa$ is the Born-Infeld constant
and $1/\kappa$ is the field strength for which non-linearities are significant. The electric field $\mathbf{E} = (E_x,E_y,E_z)$ and magnetic field $\mathbf{B} = (B_x,B_y,B_z)$ are induced from the electric $1$-form $E=E_x dx + E_y dy + E_z dz$ and magnetic $1$-form $B=B_x dx + B_y dy + B_z dz$ respectively, and it follows $X = \mathbf{E}^2-\mathbf{B}^2$, $Y = 2\mathbf{E}\cdot\mathbf{B}$.
\section{Electromagnetic Waves in a Constant Background Magnetic Field}
\subsection{Preliminaries}
\begin{figure}
\begin{center}
\begin{picture}(300,150)
 \put(20,20){\circle{8}}
 \put(20,20){\circle{1}}
 \put(20, 20){\vector(0, 1){50}}
 \put(20, 20){\vector(1, 0){50}}

 \put(17, 4){\makebox(0, 0){$z$}}
 \put(80, 18){\makebox(0, 0){$x$}}
 \put(28, 70){\makebox(0, 0){$y$}}

\put(170,60){\circle{8}}
 \put(170,60){\circle{1}}
 \put(170, 60){\vector(0, -1){50}}
 \put(170, 60){\vector(1, 0){50}}
 \put(170, 60){\vector(-1, 0){50}}

 \put(230, 70){\makebox(0, 0){$\mathbf{B}_{background}$}}
 \put(190, 30){\makebox(0, 0){$\mathbf{B}_{wave}$}}
 \put(140, 70){\makebox(0, 0){$\mathbf{E}_{wave}$}}
\end{picture}
\caption{The relative directions of the plane wave corresponding to the Faraday $2$-form (\ref{Fone}), where the wave is propagating out of the page, i.e. in the positive z-direction.}\label{Figure1}
\end{center}
\end{figure}
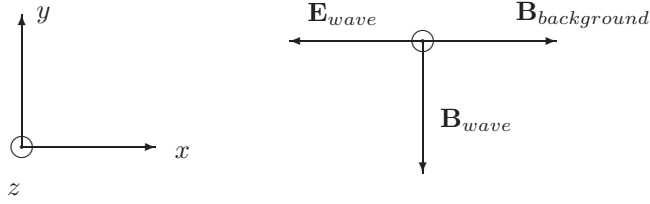

Consider a linearly polarized electromagnetic plane wave travelling through a region of constant magnetic field as per Figure \ref{Figure1}. The direction of propagation of the wave is orthogonal to the background magnetic field and the wave's electric field is parallel to the background magnetic field. The Faraday $2$-form $F$ may be written as
\begin{equation}
F = \mathcal{E}(z-vt)(dz-v\,dt)\wedge dx - B_x\, dy\wedge dz\label{Fone}
\end{equation}
by appropriately orienting the Cartesian coordinate frame so that the wave propagates along $z$ and the background magnetic field is aligned along $x$. The smooth function $ \mathcal{E}$ encodes the wave's electric and magnetic fields and $v$ is the phase velocity of the wave.

Equations (\ref{maxwell1}), (\ref{maxwell2}) and (\ref{Gdef}) may be used to show that Lagrangians of the form 
\begin{equation}
\mathcal{L}(X,Y)=c_1 + c_2Y+\mathcal{F}(X+\lambda Y^2)\label{lagfaminit}
\end{equation}
yield field equations that have (\ref{Fone}) as an {\it exact} solution. The real constant $\lambda$ is positive, $c_1$ and $c_2$ are real constants and $\mathcal{F}$ is a smooth function satisfying $\mathcal{F}(X+\lambda Y^2) \approx X/2$ to first order in $X$ and $Y$ (which ensures that the usual linear vacuum Maxwell equations are recovered in the weak-field limit). Although it is possible to simply verify (\ref{lagfaminit}) by direct substitution, it is instructive to show how one is led to this result from first principles.

Firstly, note that (\ref{Fone}) satisfies (\ref{maxwell1}) by construction, hence it is only necessary to consider (\ref{maxwell2}). Using (\ref{maxwell2}), (\ref{Gdef}) and (\ref{Fone}) one obtains
\begin{align}
\label{raw_component_1}
&(\gamma^2Bv\partial^2_{XY}\mathcal{L}\big|_P+\mathcal{E}\partial^2_X\mathcal{L}\big|_P){\cal E}^\prime=0,\\
\label{raw_component_2}
&(\partial_X\mathcal{L}\big|_P-2\mathcal{E}^2\gamma^{-2}\partial_X^2\mathcal{L}\big|_P-4vB\mathcal{E}\partial^2_{XY}\mathcal{L}\big|_P
 - 2v^2\gamma^2B^2\partial^2_Y\mathcal{L}\big|_P){\cal E}^\prime =0,
\end{align}
where ${\cal E}^\prime$ is the derivative of ${\cal E}$, $\gamma=1/\sqrt{1-v^2}$ is the Lorentz factor of the phase velocity $v$ of the wave, $B\equiv B_x$ and $P\equiv (X = -\gamma^{-2}\mathcal{E}^2-B^2, Y=-2Bv\mathcal{E})$ is a point in $(X,Y)$-space corresponding to (\ref{Fone}). Equation (\ref{Y_def}) yields $Y=-2B v\mathcal{E}$, which is used to eliminate $\mathcal{E}$ and cast (\ref{raw_component_1}) and (\ref{raw_component_2}) as
\begin{align}
\label{messyresteq1}
&\gamma^2Bv\partial^2_{XY}\mathcal{L}-\frac{Y}{2Bv}\partial^2_X\mathcal{L}\simeq 0,\\
\label{messyresteq2}
&\partial_X\mathcal{L}-\frac{Y^2}{2B^2v^2\gamma^{2}}\partial_X^2\mathcal{L}+2Y\partial^2_{XY}\mathcal{L}
 - 2v^2\gamma^2B^2\partial^2_Y\mathcal{L}
\simeq 0,
\end{align}
where the function ${\cal E}^\prime$ has been removed and $\simeq$ indicates equality on restriction to the subset ${\cal U} = \{(X,Y)\,|\, X+Y^2/(4B^2\gamma^2 v^2) - B^2 = 0\}$ of $(X,Y)$-space.

A further assumption must now be made in order to proceed; (\ref{messyresteq1}) and (\ref{messyresteq2}) are extended away from ${\cal U}$ by demanding
\begin{align}
&\gamma^2Bv\partial^2_{XY}\mathcal{L}-\frac{Y}{2Bv}\partial^2_X\mathcal{L}= 0,\label{eq_messyresteq1}\\
\label{eq_messyresteq2}
&\partial_X\mathcal{L}-\frac{Y^2}{2B^2v^2\gamma^{2}}\partial_X^2\mathcal{L}+2Y\partial^2_{XY}\mathcal{L} - 2v^2\gamma^2B^2\partial^2_Y\mathcal{L}
= 0.
\end{align}
Inserting the general solution to (\ref{eq_messyresteq1}) into (\ref{eq_messyresteq2}) yields
\begin{equation}
\mathcal{L}(X,Y)=c_1 + c_2 Y+\mathcal{F}\left(X+\frac{1}{4B^2v^2\gamma^2}Y^2\right)\label{Bxsol}
\end{equation}
where $c_1$ and $c_2$ are constants and $\mathcal{F}$ is any smooth function.

The constant $c_1$ is significant if the gravitational field is dynamical (it is the cosmological constant); however, here the spacetime metric is prescribed and $c_1$ does not place a role. Although the term $c_2 Y$ in (\ref{lagfaminit}) may play a role in spatially bounded domains, this term is not considered further in the present article.

Using (\ref{Bxsol}), it follows that theories whose Lagrangians have the form (\ref{lagfaminit}) admit the same solution (\ref{Fone}) with phase velocity $v$ satisfying
\begin{equation}
v^2 = \frac{1}{1+4\lambda B^2}.
\end{equation}
\subsection{Coplanar background magnetic field, wave vector and transverse electric field}
It is straightforward to extend the analysis given in the previous section to more general configurations in which the background magnetic field, wave vector and transverse projection of the electric field (with respect to the direction of propagation) are coplanar. In particular, extension of the magnetic field to an arbitrary constant vector in the $(x,z)$ plane is accommodated by introducing a longitudinal component to the electric field:
\begin{align}
\label{coplan_B_k_E}
&F = \mathcal{E}(z-vt)(dz-vdt)\wedge dx - B_x dy\wedge dz - B_z dx\wedge dy + \chi\mathcal{E}(z-vt)\,dt\wedge dz,
\end{align}
with $\chi$ being a real constant. The electromagnetic field (\ref{coplan_B_k_E}) has previously been shown to be an exact solution to the vacuum Born-Infeld field equations~\cite{aiello:2007}. 

Equation (\ref{raw_component_1}) arose from the $dy$ component of $\star d\star G = 0$ in the previous case. Using the same strategy in the present case (i.e. inserting the general solution to the $dy$ component of $\star d\star G=0$ into the remaining components of $\star d\star G=0$) leads directly to (\ref{lagfaminit}) with
\begin{equation}
\label{v_chi_coplanar}
v^2 = \frac{1+4\lambda B_z^2}{1+4\lambda(B_x^2+B_z^2)},\quad \chi=\frac{4\lambda B_x B_z v}{1+4\lambda B_z^2}.
\end{equation}
Theories of the form (\ref{lagfaminit}) with identical $\lambda$ possess (\ref{coplan_B_k_E}) as a solution.
\subsection{Non-coplanar background magnetic field, wave vector and transverse electric field}
No simple modification of (\ref{coplan_B_k_E}) has been found to be valid in the general case. In particular, using
\begin{align}
\label{general_case_Fone}
&F = \mathcal{E}(z-vt)(dz-vdt)\wedge dx -B_x dy\wedge dz - B_y dz \wedge dx - B_z dx\wedge dy + \chi\mathcal{E}(z-vt)\, dt\wedge dz
\end{align}
and (\ref{lagfaminit}) in (\ref{maxwell1}), (\ref{maxwell2}) leads to the condition $B_y=0$ if no additional constraints are imposed on ${\cal F}$. However, as shown in Ref. \cite{aiello:2007}, equation (\ref{general_case_Fone}) is an exact solution to the vacuum Born-Infeld equations, i.e. (\ref{maxwell1}), (\ref{maxwell2}) with ${\cal L}={\cal L}_{\rm BI}$ (see Eq. (\ref{BI_def})) and
\begin{equation}
v^2 =  \frac{1+\kappa^2 B_z^2}{1+\kappa^2(B_x^2+B_y^2+B_z^2)},\quad
\chi = \frac{\kappa^2 B_x B_z v}{1+\kappa^2 B_z^2}. 
\end{equation}
\section{Electric-magnetic duality}
Following Gibbons~\cite{gibbons:1995}, one may elevate electric-magnetic duality invariance to a fundamental property of the electromagnetic field in source-free regions. An electric-magnetic duality transformation is an endomorphism on the space of solutions of (\ref{maxwell1}), (\ref{maxwell2}) given by the $SO(2)$ action
\begin{equation}
\left(\begin{array}{c}
F\\
\star G
\end{array}
 \right)
\rightarrow
\left(\begin{array}{cc}
\cos\alpha & \sin\alpha\\
-\sin\alpha & \cos\alpha
\end{array}
 \right)
\left(\begin{array}{c}
F\\
\star G
\end{array}
 \right)
\end{equation}
where $\alpha$ is a real constant. Invariance under infinitesimal duality transformations leads to the condition 
\begin{equation}
\label{EM_symmetry}
\star(F\wedge F) - \star(G\wedge G) = C
\end{equation}
on Lagrangian-based theories~\cite{gibbons:1995}, and setting the constant $C$ to zero selects the family of electric-magnetic duality invariant theories containing $G=F$ (linear Maxwell theory).

After removing the topological term from the action corresponding to (\ref{lagfaminit}) by setting $c_2=0$, it follows immediately that the electromagnetic duality condition $F\wedge F=G \wedge G$ is satisfied if 
\begin{equation}
\mathcal{F}(\xi)=\frac{1}{4\lambda}\left(1 - \sqrt{1-4\lambda \xi}\right)
\end{equation}
where a constant of integration has been fixed by demanding $\mathcal{F}(X+\lambda Y^2) \approx X/2$ in the weak-field limit. Hence, the intersection of the $C=0$ family of electric-magnetic duality invariant Lagrangians and the $c_1=c_2=0$ family of Lagrangians is the Born-Infeld Lagrangian (\ref{BI_def}).
\section{Conclusion}
It has been shown that (\ref{coplan_B_k_E}) is a solution to a class of theories of vacuum non-linear electrodynamics, and Born-Infeld electrodynamics is singled out from that class if electric-magnetic duality invariance is invoked. However, without electric-magnetic duality invariance, non-coplanar configurations are the most appropriate choice for attempts to constrain theories of non-linear electrodynamics using time-of-flight measurements of an electromagnetic pulse propagating through a constant magnetic field.
\section{Acknowledgments}
The authors are members of the ALPHA-X consortium funded by EPSRC. They are also grateful for support provided by the Cockcroft Institute of Accelerator Science and Technology (STFC).

\end{document}